\documentclass{emulateapj}

\slugcomment{Accepted to ApJL}

\shorttitle{Thermally-emitting dust around HD~32297}
\shortauthors{Moerchen et al.}

\begin{document}

\title{12 and 18 micron images of dust surrounding HD~32297}

\author{Margaret M. Moerchen \altaffilmark{1}, Charles M. Telesco \altaffilmark{1}, James M. De Buizer \altaffilmark{2}, Christopher Packham\altaffilmark{1} \& James T. Radomski \altaffilmark{2}}

\altaffiltext{1}{University of Florida, 211 Bryant Space Science Center, P. O. Box 112055, Gainesville, FL 32611-2055; margaret@astro.ufl.edu}
\altaffiltext{2}{Gemini Observatory, Casilla 603, La Serena, Chile}

\begin{abstract}
We present the first subarcsecond-resolution images at multiple mid-IR wavelengths of the thermally-emitting dust around the A0 star HD~32297.  Our observations with T-ReCS at Gemini South reveal a nearly edge-on resolved disk at both 11.7~$\mu$m and 18.3~$\mu$m that extends $\sim$150~AU in radius.  The mid-IR is the third wavelength region in which this disk has been resolved, following coronagraphic observations by others of the source at optical and near-IR wavelengths.   The global mid-IR colors and detailed consideration of the radial color-temperature distribution imply that the central part of the disk out to $\sim$80~AU is relatively deficient in dust.
\end{abstract}


\section{Mid-IR imaging of debris disks}

Many main-sequence stars that are old enough for their primordial disks to have dissipated still emit significant excess infrared flux. This emission is usually attributed to starlight-heated dust located in a debris disk where the dust population is continually resupplied by processes such as collisions between planetesimals or sublimation of cometary material (e.g. Aumann et al. 1984, Backman et al. 1992; Wyatt et al. 1999).  Mid-IR ($\sim$8-25~$\mu$m) imaging is a useful tool for investigating these disks.  Photospheric emission at these wavelengths is often small compared to the disk emission, and so fruitful imaging of these disks, especially the innermost regions, does not require the use of a coronagraph, which blocks the starlight but also hides the key regions of interest near the star.  Large (8-10~m) ground-based telescopes can directly probe these inner subarcsecond regions of the disk at mid-IR wavelengths.  This emission can also sample a different particle population than those seen in scattered optical or near-IR light, thereby making it a valuable complementary approach; submicrometer-to-micrometer-sized particles are particularly evident in the mid-IR.

Here we report the first multi-wavelength mid-IR images that resolve the debris disk of HD~32297, and we note the independent discovery of resolved thermal emission 
by \citet{fit07}.
These observations were motivated by high-contrast coronagraphic imaging of scattered light in the extended dust disk around HD~32297 in near-IR and optical bandpasses \citep{sch05,kal05b}.  The high fractional infrared luminosity (0.0027; Schneider et al. 2005) of this A0 star indicates a considerable contribution from thermal emission that, it seemed to us, would likely be resolvable at a distance of 112~pc (Perryman et al. 1997, hereafter P97) at mid-IR wavelengths.  Over a hundred debris disk candidates have been identified photometrically (e.g., Rieke et al. 2005), but only about a dozen have been spatially resolved at any wavelength.  Of the few resolved disks, seven belong to A-type stars like HD~32297, with two of these (Vega and Fomalhaut [Su et al. 2005; Stapelfeldt et al. 2004]) having surface brightnesses too low to be detected in the infrared with reasonable integration times from the ground.  These and most other resolved debris disks around A stars ($\beta$ Pic, HR~4796A, 49~Ceti, and HD~141569) extend more than 100~AU in radius \citep{bac92, tel88,lag94,tel05,jay98,tel00,wah07,fis00,mar02}; $\zeta$~Lep remains the exception among resolved disks, being comparable in size to our asteroid belt \citep{che01, moe07}.  HD~32297 is therefore an important addition to the small but growing sample of debris disks resolved at mid-IR wavelengths.
	
\section{Observations of HD~32297}

{\it Data.}---	We obtained mid-IR images of HD~32297 on four nights in 2006 (February 3, March 4, 9, and 10 [UT]) at Gemini South (program ID: GS-2005B-DD-8) with T-ReCS (Thermal Region Camera and Spectrograph), using the narrowband Si-5 ($\lambda_c$~=~11.66~$\mu$m, $\Delta\lambda$~=~1.13~$\mu$m) and narrowband Qa ($\lambda_c$~=~18.30~$\mu$m, $\Delta\lambda$~=~1.51~$\mu$m) filters.   T-ReCS utilizes a Raytheon Si:As BIB detector with 320 x 240 pixels.  Each pixel subtends 0.09", and the total field of view is 28.8" x 21.6".  The observations used the standard mid-IR technique of chopping and nodding to remove time-variable sky background, telescope thermal emission, and low-frequency detector noise; the chop throw was 15".  The data were reduced with the Gemini IRAF package.  To account for different exposure times, sky transparency, and sky background, the images from the four nights were weighted by the spatially integrated signal-to-noise ratio to produce the final images from each filter.  The flux standard HD~37160 was observed before and after the group of target observations; this source also served as a point-spread-function (PSF) comparison star.  HD~37160 is a G star located 8.7$\rm^o$ away from HD~32297 and is approximately 125 times brighter than HD~32297 at 11.7~$\mu$m and 25 times brighter at 18.3~$\mu$m.  For the four nights of observations, the PSF FWHM ranged from 0.37''--0.46'' at 11.7~$\mu$m and from 0.53''--0.60'' at 18.3~$\mu$m.  By quadratic subtraction, we infer that the greatest deviation from the $\lambda$/D diffraction limit (0.31'' at 11.7~$\mu$m and 0.49'' at 18.3~$\mu$m) occurred with seeing disks of 0.33'' at 11.7~$\mu$m and 0.34'' at 18.3~$\mu$m.  The final PSF images were combined with the same weights as the source images and indicated a resolution (FWHM) of 0.40'' at 11.7~$\mu$m and 0.54'' at 18.3~$\mu$m.  The total (disk + star) flux densities of HD~32297 are 53~mJy at 11.7~$\mu$m and 90~mJy at 18.3~$\mu$m.  Observations of the flux standards were not repeated throughout the night, so we adopt nominal calibration uncertainties of 10\% at 11.7~$\mu$m and 15\% at 18.3~$\mu$m, which are typical for photometric variations in the mid-IR (e.g. De Buizer et al. 2005, Packham et al. 2005).   Our 11.7~$\mu$m flux density is consistent with the $IRAS$ 12~$\mu$m upper limit.  $IRAS$ detected HD~32297 (IRAS 04597+0723) at 25~$\mu$m (moderate quality) and~60~$\mu$m (high quality).

\begin{figure*}
\epsscale{1}
\plotone{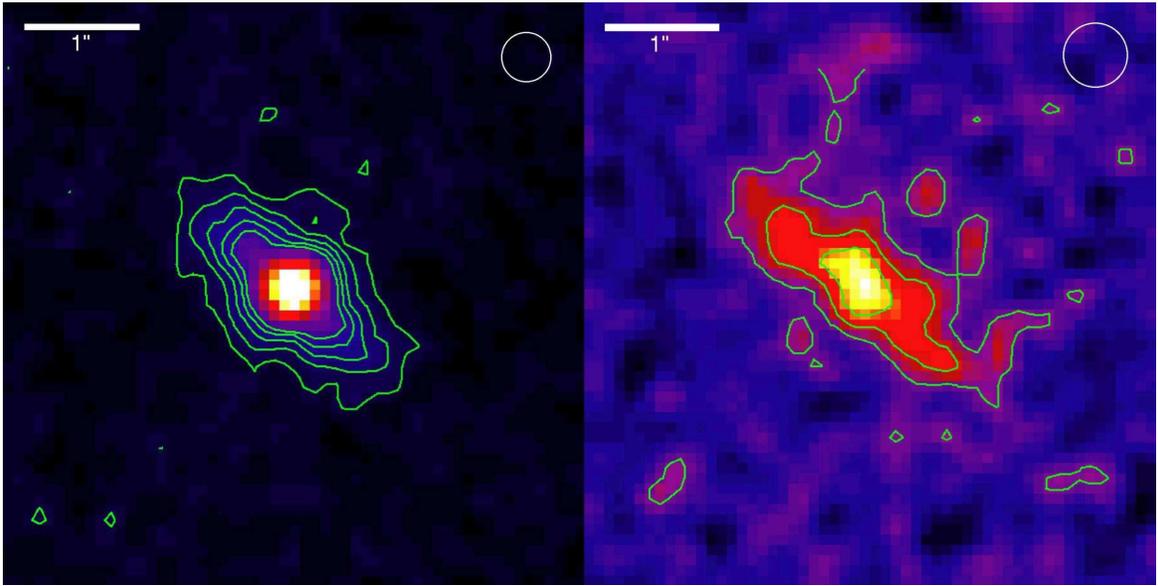}
\caption{11.7 $\mu$m (left) and 18.3~$\mu$m (right) images of HD~32297.  Contours are linearly spaced at 3, 6, 9, 12, and 15 times the 1-$\sigma$ noise level (mJy pixel$^{-1}$) in the 11.7~$\mu$m image, and at 3, 6, and 9 times the 1-$\sigma$ noise level (mJy pixel$^{-1}$) in the 18.3~$\mu$m image.  The 11.7~$\mu$m and 18.3~$\mu$m images are gaussian-smoothed by 2 and 3 pixels, respectively.  The circle at the upper right in each image corresponds to the size of the FWHM of the PSF at each wavelength.\label{fig:images} }
\end{figure*}

{\it Disk geometry and morphology.}--- We observe an apparent nearly edge-on disk (Figure \ref{fig:images}) with a total extent (at the 3-$\sigma$ level) of 291~AU (2.58'') and 341~AU (3.02'') at 11.7 and 18.3~$\mu$m, respectively.  With ellipse fitting of isophotes in the 11.7~$\mu$m image, we find a PA for the major axis of 51.1$\rm^o$~$\pm$~2.4$\rm^o$ (east of north), consistent to within 2$\rm^o$ with the value measured for the entire disk in near-IR images \citep{sch05} and for the southwest side of the disk in optical images \citep{kal05b}.  The mid-IR spatial extent is only half that measured at near-IR wavelengths \citep{sch05}. 
			
	At 11.7~$\mu$m, the integrated flux of the NE side of the disk is 1.34~$\pm$~0.09 times as bright as the SW side; the statistical significance of the difference in the brightness between the two sides is 6.1 $\sigma$.  The fluxes used in this calculation measure the total flux on each side exterior to the 0.75'' region corresponding to the first Airy ring, in order to exclude most of the photospheric flux.  This asymmetry is opposite to that determined by \citet{sch05}, where the scattered light in the coronagraphic image is brighter in the SW by a difference of 2.6 $\sigma$.  At 18.3~$\mu$m, the SW side is 1.28~$\pm$~0.29 times as bright as the NW side (opposite to that at 11.7~$\mu$m), but this difference is not significant (1.5 $\sigma$).  The fluxes for each side at 18.3~$\mu$m are measured from the central star position outward, since the photospheric contribution at this wavelength is small (see Table \ref{tab:fluxes}).  Other than the modest difference in brightness between sides, there are no obvious morphological features in the disk at either wavelength.  Asymmetries are seen in many disks, and they may have diverse origins such as resonant trapping or cataclysmic collisions \citep{kal05a,tel05}.  However, until deeper multi-wavelength images are available, detailed considerations of any asymmetries in HD~32297 seem to us to be unwarranted.

\section{Discussion}

{\it Comparison with key debris disk archetypes.}---  We compare the global mid-IR characteristics of HD~32297 to those of two important archetypes for disks around A stars that are probably of comparable ($\sim10^{7}$~Myr) age: $\beta$~Pic and HR~4796A.  HD~32297's age is currently not well constrained, but a possible association of the source with the Gould Belt or Taurus-Aurigae supports an age estimate of $\sim$30~Myr.  However, \citet{kal05b} notes that the optically observed morphology could be interpreted as the result of outflows suggestive of a much younger age.  The fractional IR luminosity of HD~32297 (0.0027; Schneider et al. 2005) is similar to those of $\beta$~Pic and HR~4796A: 0.0015 \citep{dec03}, and 0.005 \citep{sch99}, respectively.  HD~32297 is much farther away (112~pc [P97]) than $\beta$~Pic (19~pc [P97]) and HR~4796A (67~pc [P97]), which results in a lower total signal-to-noise ratio and poorer resolution in its images.  Therefore, it is not possible with our current images to study the inner region of the disk closest to HD~32297 ($\lesssim$30~AU) in as great detail as these other two sources.  We can, however, usefully compare for the first time its mid-IR colors to those of the two disks. 

We estimate the photospheric flux densities for HD~32297 at 11.7~$\mu$m and 18.3~$\mu$m by extrapolating the K-band magnitude of 7.59 \citep {cut03} to 10~$\mu$m.  We assume that the flux density varies as $\nu^{1.88}$ over this wavelength range, as is estimated by \citet{kur79} to be appropriate for an A0 star (e.g. Jura et al. 1998).  Beyond 10~$\mu$m, we adopted the standard Rayleigh-Jeans ($\nu^{2}$) relation to estimate photospheric flux densities and resulting excess flux densities as given in Table 1.  The 24.5~$\mu$m flux density of 198~$\pm$~32~mJy was interpolated from our 11.7~$\mu$m measurement and the $IRAS$ detection at 25~$\mu$m.

\begin{figure}
\includegraphics[width=\columnwidth]{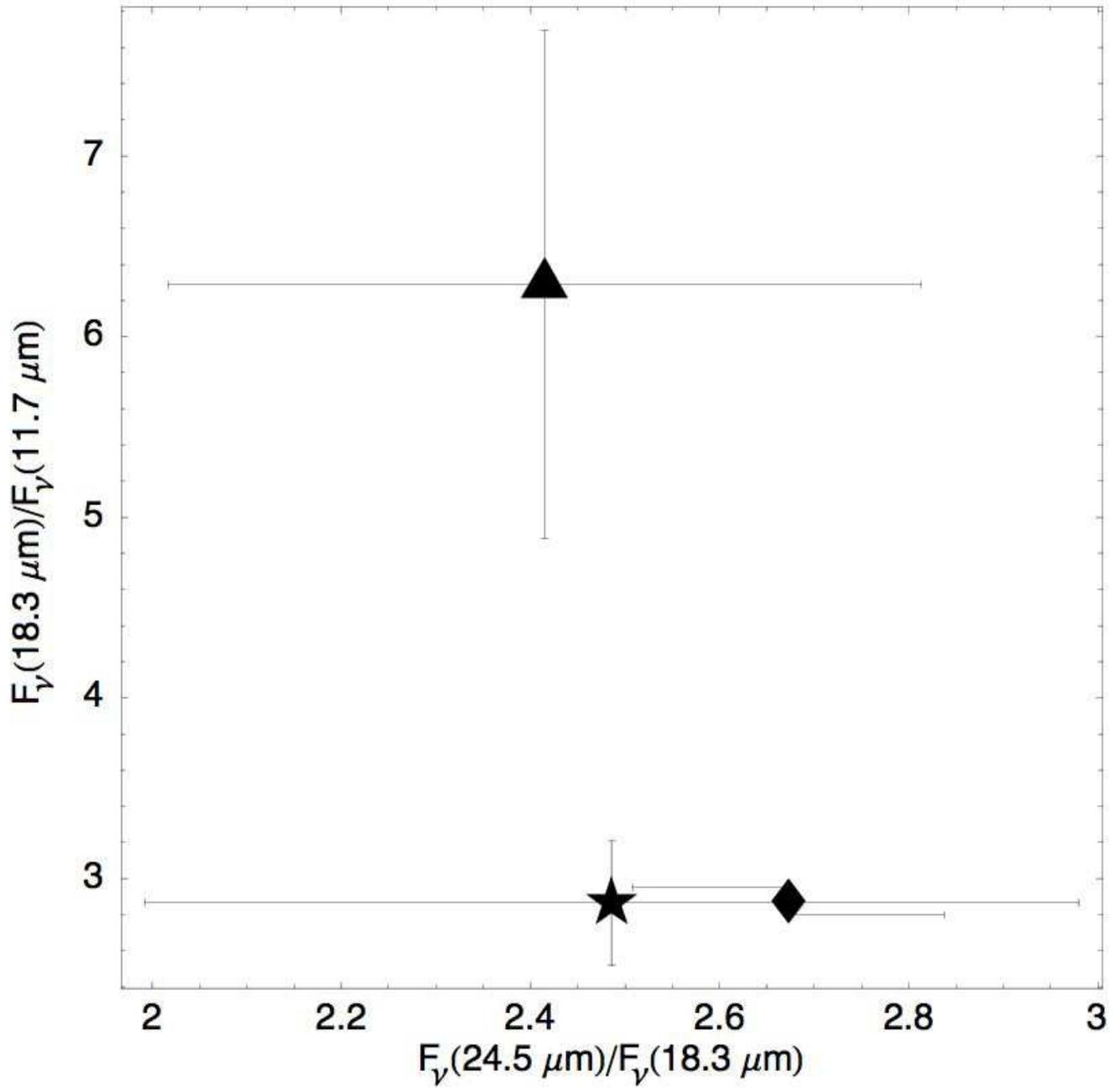}
\caption{Color-color plot of mid-IR fluxes of HD~32297 (star symbol) and the two archetype disks, $\beta$~Pic (diamond) and HR~4796A (triangle).  Fluxes and uncertainties for HD~32297 are in Table \ref{tab:fluxes}. \label{fig:colors}}
\end{figure}

The color-color plot (Figure \ref{fig:colors}) of HD~32297, $\beta$~Pic and HR~4796A shows that all three sources possess similar $F_{\nu}$(24.5~$\mu$m)/$F_{\nu}$(18.3~$\mu$m) colors, but the $F_{\nu}$(18.3~$\mu$m)/$F_{\nu}$(11.7~$\mu$m) color of HR~4796A is twice that of both $\beta$~Pic and HD~32297 \citep{tel05,tel00,wah05}.  The high value for this latter color for HR~4796A is likely due to a deficiency of hot 11.7~$\mu$m-emitting dust very close to the star, as is indeed seen in mid-IR images of the source (e.g. Jayawardhana et al. 1998; Koerner et al. 1998; Telesco et al. 2000) where the region interior to 70~AU is largely devoid of dust.  The $F_{\nu}$(18.3~$\mu$m)/$F_{\nu}$(11.7~$\mu$m) color of HD~32297 appears more consistent with its innermost particles having smaller average radial distances, more comparable to the case of $\beta$~Pic. 	In $\S$3.2, we provide additional evidence and discussion regarding the detailed geometry of the central clearing.

{\it Characteristic particle temperatures.}--- The mid-IR color temperature computed from the two excess flux densities is 186~$\pm$~15~K.  The uncertainties in the temperature reflect those from measurement and photometric calibration.  If the dust particles were blackbody emitters, the color temperature would imply a distance of $\sim$7~AU for the majority of the particles.  However, since we observe the disk to extend at least ten times farther, most of the constituent particles must be small, inefficient emitters that are less than a few microns in radius for them to be so warm \citep{prz03}.   For comparison, we also compute the temperature of the dust particles assuming an emission efficiency proportional to frequency, which yields 156~$\pm$~10~K.

\begin{figure}
\includegraphics[width=\columnwidth]{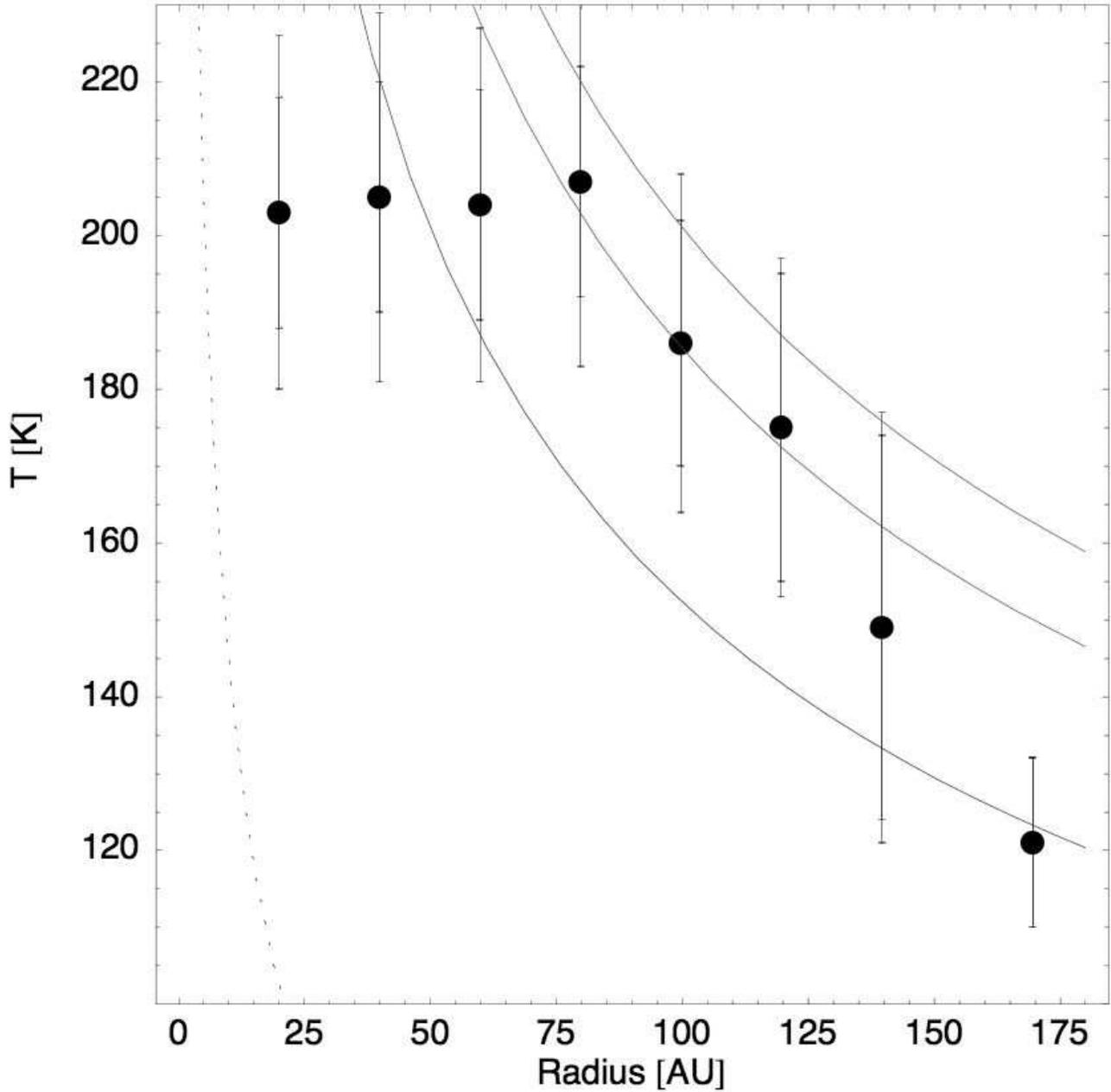}
\caption{Radial color temperature profile of HD~32297, as derived from 11.7 and 18.3~$\mu$m flux density measurements along the disk.  Large error bars reflect the combined photometric and measurement uncertainties and small error bars reflect only the measurement uncertainties.  The dotted line indicates the expected blackbody temperature change with distance from the star and the three solid lines indicate predicted temperatures (as discussed in \citet{bac93} and \S3.2) for characteristic grain sizes: 0.25~$\mu$m (left), 0.075~$\mu$m (center), and 0.05~$\mu$m (right). \label{fig:temps}}
\end{figure}

We measured the radial brightness profiles of the photosphere-subtracted disk by summing over a 0.8'' swath along the disk axis and then summing the two sides of the disk.  The radial bin widths are $\sim$20~AU, with the exception of the outermost data point that reflects the mean value for a $\sim$40~AU bin to increase S/N.  For each filter, the PSF was scaled to the level of the photosphere (see Table \ref{tab:fluxes}) and then subtracted from the disk image.  To achieve the same resolution in the two images, each image was then convolved with a gaussian profile having a width corresponding to the PSF FWHM in the other filter.  We used these brightness profiles to generate a color temperature profile of the disk, shown in Figure \ref{fig:temps}.  We see that the color temperature remains constant to within a few kelvins from the center of the disk out to $\sim$80~AU.  Particle temperature should decrease as distance from the star increases.  Therefore, this zone of constant temperature implies that the particles dominating the emission are only viewed in projection and are, in fact, at approximately the same physical distance from the star.  Based on the location where the temperature begins to decline, that distance is 80~AU, which must correspond to the approximate outer boundary of a central region that is highly deficient in dust, as suggested by the color-color plot in $\S$3.1.  However, in apparent contradiction to Figure \ref{fig:colors}, in which the colors of HD~32297 are nearly the same as those of $\beta$~Pic, this inferred central clearing size is almost identical to that determined for HR~4796A (e.g. Jayawardhana et al. 1998; Koerner et al. 1998; Telesco et al. 2000).  We suggest that both findings can be reconciled in a picture where the zone markedly deficient in dust is approximately the same size as the one in HR~4796A, but that this zone still contains significantly more dust than the cleared region of HR~4796A, which is almost completely free of debris.  Thus we may be observing a debris disk with a morphology intermediate between $\beta$~Pic and HR~4796A.  We do not mean to suggest that these stars correspond to a temporal sequence, since that is inconsistent with the known ages of $\beta$~Pic and HR~4796A.  However, the disk in HD 32297 appears to broaden our appreciation of the range of morphologies that can occur during disk evolution.

We plot temperature curves in Figure \ref{fig:temps} for a blackbody and for particles of three characteristic radii (0.05~$\mu$m, 0.075~$\mu$m, and 0.2~$\mu$m).  These latter curves were calculated using equations presented in \citet{bac93}, which enable a particle temperature estimate at a given distance from the star for particles with a range of characteristics, described by a critical wavelength parameter $\lambda_{o}$.  This parameter is related to radiation efficiency, such that the efficiency is assumed to be nearly unity for wavelengths shorter than $\lambda_{o}$.  The value of $\lambda_{o}$  depends on composition, and we adopt the case where $\lambda_{o}$ is approximately equal to the particle size, as expected for particles that absorb efficiently but emit inefficiently, such as graphite and amorphous silicate \citep{bac93}.  We find that HD~32297's color temperature profile does not coincide with that expected for particles of only one characteristic size.  The profile is well approximated by the curve for 0.075-$\mu$m particles between $\sim$80--120~AU, but at larger distances the color temperature falls below this curve.  Since $\lambda_{o}$ reflects both particle size and composition, a change in either (or both) of these characteristics of the dominant population at $\sim$120~AU could produce the deviation we observe from the temperature curve for a single $\lambda_{o}$ value.  

Such a radial transition might occur at the snow line.  The snow line in protoplanetary disks is expected to lie where temperatures fall to $\sim$145--170~K (e.g. Hayashi 1981; Sasselov \& Lecar 2000).  We estimate color temperatures below this level for the two outermost points in the disk, beyond $\sim$140~AU.  An increase in particle surface density beyond the putative snow line would likely result in the growth of larger icy particles than would be found in the inner region of the disk.  Such particles would show different absorption and emission behavior compared to non-ice-enhanced particles closer to the star, not only because of their larger size, but also because of the significant change in composition, which dictates the particle heating and cooling processes (e.g. Podolak \& Zucker 2004).  This idea is purely speculative and must be investigated further with spatially resolved spectroscopic observations and subsequent modeling.

While the disk of HD~32297 must be explored with deeper direct imaging and spectroscopy to clarify the distribution and composition of dust, we have established the key conclusion that HD~32297 is yet another debris disk with a central clearing.  The number of disks currently resolved with imaging is still small, but most of these resolved sources share the common feature of a cleared central area.  By studying these debris disks further, we can better understand how their inner regions are shaped by planets or other processes (e.g., Klahr \& Lin 2001; Kalas et al. 2005).

\acknowledgments

We wish to acknowledge very useful comments from the referee, Karl Stapelfeldt.  MM gratefully acknowledges fellowship support from the Michelson Science Center.  Observations were obtained at the Gemini Observatory, operated by AURA, Inc., under agreement with the NSF on behalf of the Gemini partnership. 

{\it Facilities:}  \facility{Gemini:South (T-ReCS)}.

\clearpage

\begin{deluxetable}{lccccccccccc}
\tabletypesize{\scriptsize}
\tablecaption{Flux Densities from Ground-Based Observations\label{tab:fluxes}}
\tablehead{
\colhead{Flux density (mJy)} &
\colhead{11.7 $\mu$m} &
\colhead{18.3 $\mu$m} &
}
\startdata
Total \tablenotemark{a}   & $53$ & $90$ &\\
Estimated photospheric\tablenotemark{b}   & $25$ & $10$\\
Excess   & $28$ & $80$\\
\enddata

\tablenotetext{a}{We assume that photometric variations dominate the uncertainties, and we adopt conservative estimates of 10\% photometric error at 11.7~$\mu$m and 15\% at 18.3~$\mu$m (see \S2.1).} 

\tablenotetext{b}{Photospheric flux density estimates are extrapolated from $K$-band photometry \citep{cut03}.  }

\end{deluxetable}

\clearpage
\end{document}